\def\BibTeX{{\rm B\kern-.05em{\sc i\kern-.025em b}\kern-.08em
    T\kern-.1667em\lower.7ex\hbox{E}\kern-.125emX}}
\documentclass[conference]{IEEEtran}
\usepackage[utf8]{inputenc}
\usepackage{tikz} 
\usetikzlibrary{positioning}
\usepackage{cite}
\usepackage{balance}
\usepackage{booktabs}  
\usepackage{amsmath,amssymb,amsfonts}
\usepackage{algorithmic}
\usepackage{graphicx}
\usepackage{textcomp}
\usepackage{xcolor}

\title{smallNet: Implementation of a convolutional layer in tiny FPGAs}
\author{
    \IEEEauthorblockN{Fernanda Zapata Bascuñán}
    \IEEEauthorblockA{\textit{Universidad Nacional de San Martín}\\
    Buenos Aires, Argentina \\
    fzapatabascuna@unsam.edu.ar}
\and
    \IEEEauthorblockN{Alan Ezequiel Fuster}
    \IEEEauthorblockA{\textit{Comisión Nacional de Energía Atómica}\\
    Buenos Aires, Argentina \\
    alan.fuster@iteda.gob.ar}
}
\date{\today}

\begin{document}
\balance
\maketitle

\begin{abstract}
Since current neural network development systems in Xilinx and VLSI require co-development with Python libraries, the first stage of a convolutional network has been implemented by developing a convolutional layer entirely in Verilog. This hand-coded design, free of IP cores and based on a filter–polynomial-like structure, enables straightforward deployment not only on low-cost FPGAs but also on SoMs, SoCs, and ASICs. We analyze the limitations of numerical representations and compare our implemented architecture, smallNet, with its computer-based counterpart, demonstrating a 5.1× speedup, over 81\% classification accuracy, and a total power consumption of just 1.5 W. The algorithm is validated on a single-core Cora Z7, demonstrating its feasibility for real-time, resource-constrained embedded applications.
\end{abstract}

\section{Introduction}
With the advancement of image processing technologies and the growing integration of control systems, a new paradigm has emerged—one that increasingly relies on processing large volumes of image data for both training and inference. This shift has underscored the need for efficient algorithm implementations, especially as hardware platforms become more accessible and ubiquitous.

The rapid rise of machine learning, particularly deep learning techniques, has further accelerated innovation in image processing. However, deploying these algorithms on embedded systems remains a significant challenge due to their high computational and memory demands \cite{zynqnet}. Most existing implementations rely on high-performance, power-intensive hardware, which makes them accessible primarily to well-funded institutions or organizations, and limits their applicability in resource-constrained or cost-sensitive environments.

In this context, the present focuses on designing a lightweight convolutional neural network entirely hand-coded in Verilog, deployable on low-cost FPGAs and free of IP cores. Unlike solutions that rely on pre-built IP cores or high-level libraries, each functional block in this design is developed from scratch, with an optimized pipeline that achieves a significant speedup compared to CPU execution. These characteristics enable the deployment of neural networks in embedded, low-budget environments—where CPU- or GPU-based alternatives are impractical due to cost and energy constraints—or serve as a first step toward ASIC implementation. 

The main objective of this work is to design and implement a lightweight convolutional neural network (CNN) optimized for real-time image classification, whose filter–polynomial-like structure allows deployment on any digital system, from low-cost FPGAs to SoMs, SoCs, and ASICs.

As goals and tasks toward the main objective, we consider:
\begin{itemize}
    \item To reduce computational complexity by simplifying the network architecture, leveraging techniques such as max pooling to downsample feature maps and decrease processing overhead.
    \item To optimize hardware resource usage to ensure compatibility with resource-constrained, low-power FPGA devices while maintaining acceptable levels of classification accuracy.
    \item To demonstrate real-time performance by implementing and validating the design in a practical embedded environment.
    \item To promote accessibility by providing a cost-effective solution suitable for deployment in environments with limited computational or financial resources.
\end{itemize}

This work is organized as a pipeline, with the following structure: Section II reviews related work and prior approaches to implementing CNNs on embedded systems. Section III introduces the proposed architecture, including key design considerations and optimization strategies for low-cost FPGA platforms. Section IV presents the experimental setup and discusses the results in terms of accuracy, performance, and resource usage. Finally, Section V concludes the paper and suggests directions for future research.

\section{Background}
A CNN is a type of deep neural network composed of cascaded convolutional layers, pooling layers, and fully connected layers. During the feed-forward phase, each convolutional layer applies a set of kernel functions—weight matrices of dimension \(k \times k\)—which slide over the input image with a certain stride. Each kernel convolves with a corresponding \(k \times k\) window of the input to produce one pixel in the subsequent feature map. This operation extracts spatially localized features by applying learned filters across the input \cite{yang2021}.
\begin{figure}[htbp]
\centering
\includegraphics[width=0.3\textwidth]{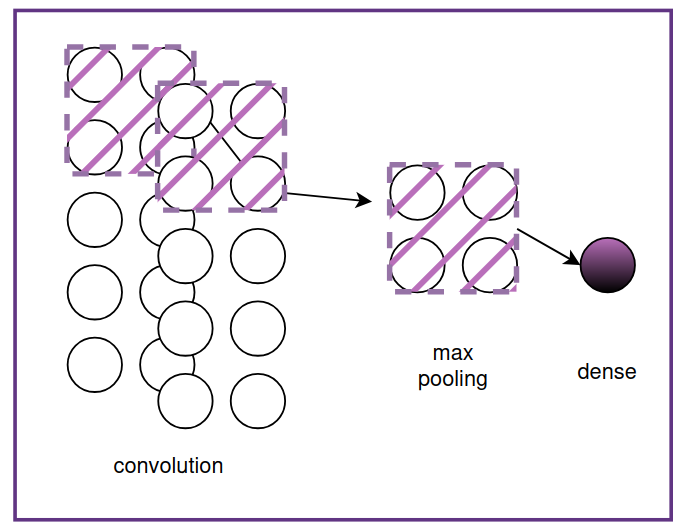}
\caption{A CNN architecture that adds convolutional layers and pooling layers before dense layers \cite{author2020}.}
\label{fig}
\end{figure}
Pooling layers perform downsampling by aggregating values within a receptive field into a single output, effectively reducing the spatial dimension of the feature maps. Similar to convolutional layers, the pooling window shifts across the input by a stride value. This process enhances spatial invariance by combining features and disregarding minor distortions or shifts in the input data. Moreover, pooling reduces the dimensionality and computational complexity for subsequent layers, making it crucial for efficient CNN architectures in resource-constrained environments \cite{zquiero2019}.
Figure 1 demonstrates the effect of dimensionality reduction achieved through max-pooling layers. While convolutional layers inherently contribute to spatial reduction, the use of max-pooling serves as a key strategy to further decrease the number of parameters and computations required, significantly improving the efficiency of the architecture.

\section{Methodology}
\subsection{Design of Neural Network}
The neural network implemented for this work follows a lightweight convolutional architecture, optimized for deployment on hardware-constrained platforms. The model is structured using the Keras Sequential API and is composed of two convolutional layers followed by pooling, a flattening operation, and a final dense output layer. All activation functions used are sigmoid, which facilitates later translation to hardware primitives compatible with FPGA logic.

\begin{figure}[htbp]
\centering
\includegraphics[width=0.15\textwidth]{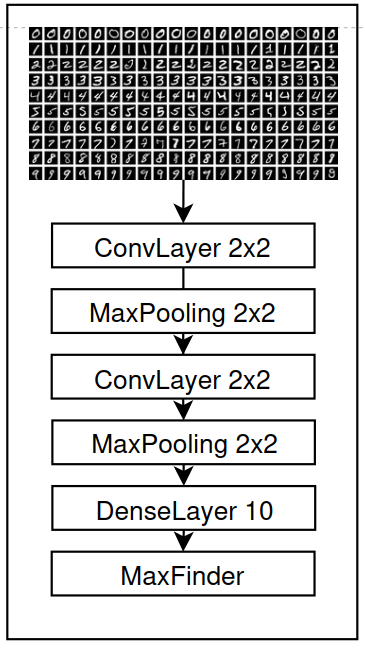}
\caption{Architecture of \textit{smallNet}, a lightweight convolutional neural network with 550 trainable parameters.}

\label{fig}
\end{figure}
The architecture is as follows:

\begin{itemize}
\item A first convolutional layer with 1 filter of size $2 \times 2$, stride 1, and \texttt{same} padding, using the sigmoid activation function. The input shape is $(28 \times 28 \times 1)$, corresponding to grayscale MNIST images.
\item A $2 \times 2$ max pooling layer, which reduces the spatial resolution by a factor of 2.
\item A second convolutional layer, also with 1 filter of size $2 \times 2$, sigmoid activation, and \texttt{same} padding.
\item A second $2 \times 2$ max pooling layer.
\item A flattening layer to convert the output tensor into a vector.
\item A fully connected dense layer with 10 output neurons and sigmoid activation, corresponding to the 10 classes of the MNIST dataset.
\item A MaxPooled 1D, also known as Max Finder, module to convert the one-hot encoded output into a binary class representation.
\end{itemize}
Figure 2 illustrates the architecture outlined in the preceding paragraph.

The model is compiled using categorical crossentropy loss function, which are suitable choices for multiclass classification tasks with one-hot encoded labels. A summary of the model confirms a very low parameter count—no more than 510 trainable parameters—which translates to approximately 1.99 KB of weight data. This compact size is ideal for FPGA-based deployment, where efficient use of memory and computational resources is essential.
The smallNet architecture was trained using the Adam optimizer with a batch size of 64 over 8 epochs. 

Figure 3 shows the training and validation loss curves obtained during the learning process, illustrating the performance of the previously described model.

\begin{figure}[htbp]
\includegraphics[width=0.5\textwidth]{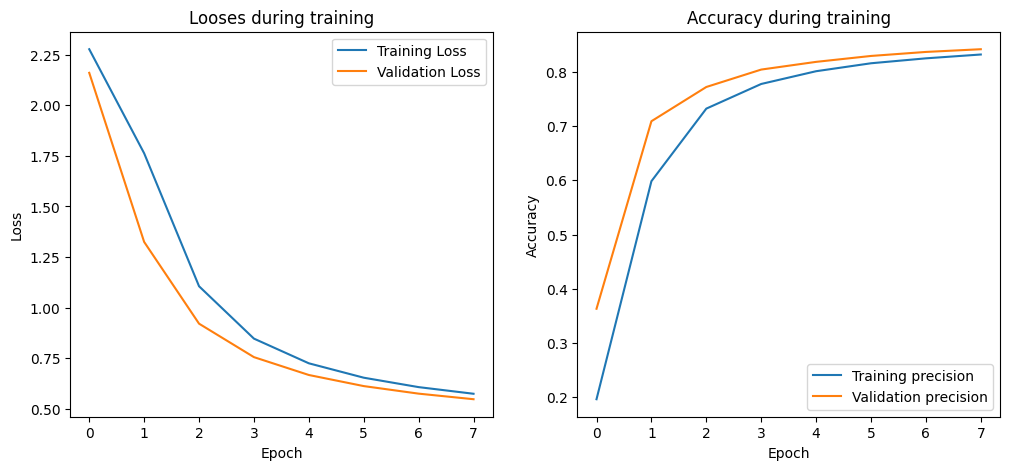}
\caption{Validation of loss and accuracy during the training of smallNet in the Keras environment.}

\label{fig}
\end{figure}

\subsection{Hardcoding layers}
The hardware design adopts a modular approach, inspired by filtering techniques commonly used in digital signal processing and widely implemented in FPGAs \cite{author2020}. Building on this foundation, the architecture replicates the structure of smallNet, consisting of convolutional layers, ReLU activations, max-pooling operations, and a fully connected output layer with sigmoid activation. Each functional block is implemented as an independent Verilog module, facilitating pipelined execution and promoting modularity and reuse throughout the design.\\
The convolutional layers support parameterized input dimensions and kernel sizes, and employ optimized multiply-accumulate (MAC) units tailored for FPGA resource efficiency. Intermediate feature maps are stored in block RAM (BRAM) and registers, facilitating efficient data flow between layers. The final layer outputs a vector of class scores, from which the predicted class (digit) is selected.

Weights and biases for each layer were extracted from a trained Keras implementation of the SmallNet model. These parameters were then converted into their 2's complement binary equivalents and hardcoded into the hardware. Based on previous studies \cite{presicion}, a fixed-point 32-bit representation was chosen, aligning with the native word size of the Zynq architecture and ensuring sufficient numerical precision during inference.\\
Figure 4 shows the implemented convolutional layer.

\begin{figure}[htbp]
\includegraphics[width=0.5\textwidth]{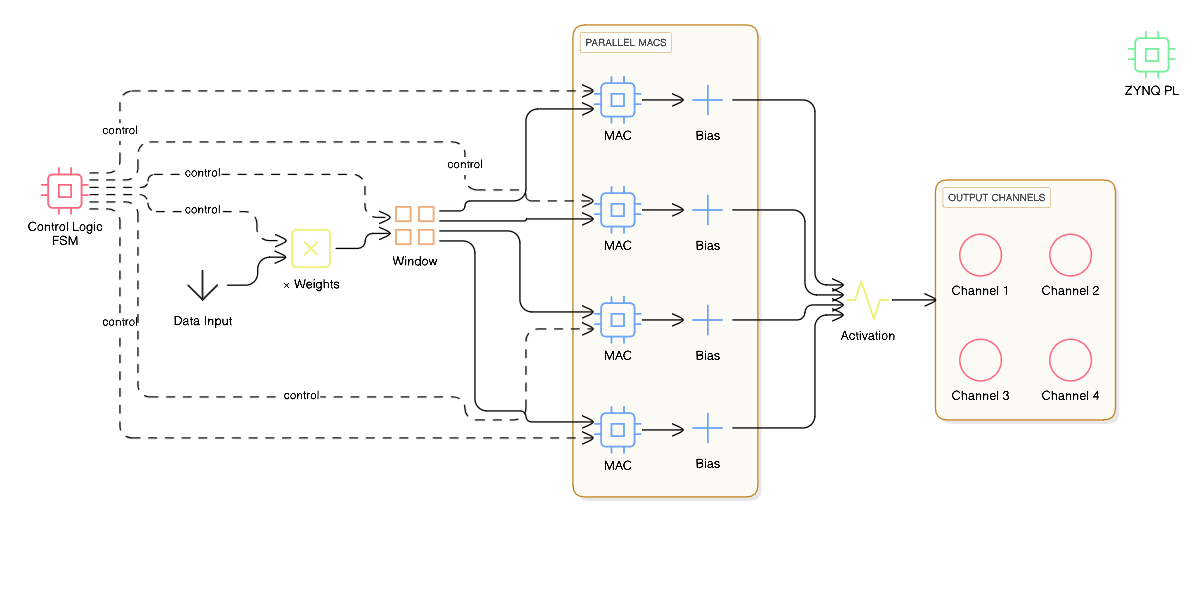}
\caption{ Hardware implementation of the convolutional neuron in Verilog. The design includes a windowing module, parallel multiply-accumulate (MAC) units with bias addition, followed by an activation function. The control logic FSM manages data flow and synchronization across the pipeline, including the application of the windowing operation.}

\label{fig}
\end{figure}
\subsection{Integration with microprocessor}
The convolution layer validation hardware was implemented on the Digilent Cora Z7 \cite{cora}, a compact and cost-effective development board based on the Xilinx Zynq-7000 All-Programmable SoC. This SoC combines a dual-core ARM Cortex-A9 processor (667 MHz) with a 7-series FPGA fabric, enabling efficient hardware/software co-design. Our solution leverages this architecture in a bare-metal environment, where no operating system is used, to maximize performance and minimize latency.

A Verilog-based neural network accelerator was developed as a custom IP core and integrated into the Zynq platform using Vivado and Vitis. The accelerator interfaces with the Zynq Processing System (PS) via AXI interconnects, with an AXI-DMA module handling high-speed data transfers between the PS and the programmable logic (PL). This allows input image data to be streamed into the accelerator and the classification results to be sent back efficiently.

The PL design contains the core neural network computation blocks along with supporting logic, including GPIO modules for control signaling and interrupt generation. On the software side, the ARM Cortex-A9 handles tasks such as configuring the DMA, managing memory, and coordinating inference. Input images are written to a dedicated memory buffer and explicitly flushed from the data cache before transmission. Once processing is complete, the PL raises an interrupt, prompting the PS to retrieve the prediction via a GPIO interface.

Figure 5 presents a block diagram of this design—simple in structure yet powerful in functionality.

The convolutional layer is fully implemented in the FPGA fabric, operating independently of the processor. The processor is used solely to load the input image into the system.

\subsection{Simulation and Testing}

To ensure correct functionality of the Verilog-based neural network design, both simulation and hardware-level testing were performed.

\textbf{Simulation:} Initial verification was conducted using behavioral simulations in Vivado. Testbenches were written to provide known input vectors and compare the output against expected values. Each neuron module was individually simulated to verify activation logic, convolutional operations, and parameter loading (weights and biases). Special attention was paid to fixed-point arithmetic handling and data alignment during kernel operations.

Waveform analysis tools were used to track signals such as valid\_in, valid\_out, and output activations to confirm proper pipeline behavior and synchronization across layers. Edge cases, such as zero-padding and max value saturation, were also included in simulation scenarios.

\textbf{Hardware Testing:} After synthesis and implementation, the design was deployed to the Cora. A C application running on the PS was used to send input images to the FPGA via AXI-DMA and receive classification results through GPIO. The input images were preprocessed from the MNIST dataset and stored in memory for transfer.

An interrupt-based mechanism was used to detect inference completion, allowing real-time verification of outputs. The system was tested with a variety of digit images to validate prediction accuracy and latency. Results were cross-checked against a software model of the neural network for consistency.

This hybrid testing approach ensured both functional correctness and hardware reliability, and allowed early detection of integration bugs or timing issues in the PL design.

\section{Results}

\subsection{Performance Metrics}

The synthesized design’s hardware resource utilization was reported as follows:

\begin{itemize}
    \item \textbf{LUTs used: } \texttt{2052} (out of \texttt{14400} available)
    \item \textbf{Flip-Flops:} \texttt{1587}
    \item \textbf{BRAMs:} \texttt{25KB}
    \item \textbf{DSP slices:} \texttt{48}
\end{itemize}

The estimated static and dynamic power consumption obtained from Vivado’s power analysis tool were:

\begin{itemize}
    \item \textbf{Total On-Chip Power:} \texttt{1.505} W
    \item \textbf{Dynamic Power:} \texttt{1.373} W
    \item \textbf{Static Power:} \texttt{113} mW
\end{itemize}

This low power consumption makes the hardware design suitable for battery-powered systems. In addition to performance, the design offers significantly lower energy consumption per inference due to pipelined parallel computation and lack of OS overhead.
\subsection{Comparison with Software Implementation}

To evaluate the benefits of hardware acceleration, the same neural network was implemented and tested using a Python (TensorFlow/Keras) model running on a standard PS side.

\begin{itemize}
    \item \textbf{Software inference time (avg):} \texttt{560 ms}
    \item \textbf{Hardware inference time (avg):} \texttt{109 ms}
    \item \textbf{Speedup: } \texttt{~5.1$\times$}
\end{itemize}

\begin{figure}
    \centering
    \includegraphics[width=0.8\linewidth]{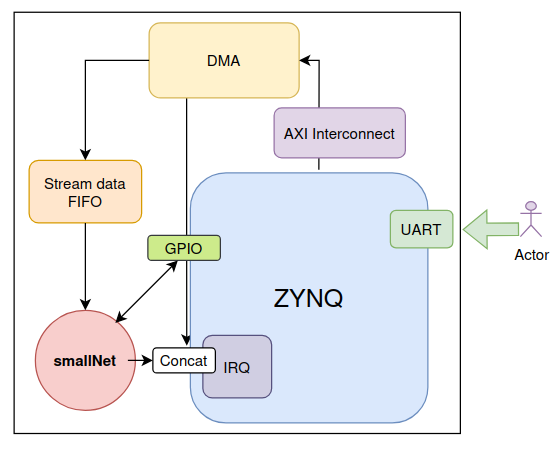}
    \caption{System-level architecture integrating \textit{smallNet}, a lightweight convolutional neural network deployed on a Zynq SoC. Data is streamed through a FIFO buffer and processed by the neural network, with GPIOs and interrupts coordinating control between programmable logic and the processing system. Results are sent via UART.}

    \label{fig:placeholder}
\end{figure}

\subsection{Visuals and Validation Results}

\textbf{Simulation Accuracy:} Using post-synthesis simulation and fixed-point test vectors from a reference software model, the classification accuracy was measured to be approximately 88.03\% on a subset of the MNIST dataset.
As a counterpart, accuracy in CPU is 93.47\%.

\textbf{Hardware Validation:} A representative set of grayscale MNIST images (scaled to 28$\times$28 pixels) was transferred to the FPGA via DMA. The output digit was read through a 4-bit GPIO interface and matched against ground truth labels.

\textbf{Hardware Accuracy:} The classification accuracy on hardware was estimated to be 81\% based on \texttt{23} test images sent via DMA.

\section{Conclusion}
This work presented an efficient hardware implementation of a basic convolutional neural network, optimized for low power consumption and high processing speed through a fully hand-coded Verilog design on the Zynq platform. The system was validated through real-time image classification on the MNIST dataset, using DMA transfers for input loading.

The results demonstrate that accurate image classification is achievable with limited resources and without requiring a high-performance CPU, supporting the feasibility of deploying intelligent embedded systems in energy-constrained environments such as IoT devices and autonomous platforms. A fully hand-coded Verilog CNN on a low-cost FPGA can deliver substantial performance gains—reducing inference latency by approximately 80\% compared to CPU execution—while maintaining power consumption levels suitable for battery-powered devices. At a hardware cost of approximately USD 149, the proposed solution offers an excellent cost-performance ratio and can be effectively deployed in IoT, lightweight robotics, and other resource-constrained applications. Moreover, the design represents a solid foundation for future ASIC implementation and, being purely algorithmic in nature, can be readily integrated into diverse digital systems to perform discrimination tasks with high efficiency.\\

\end{document}